\documentclass[utf8]{FrontiersinHarvard} 
\usepackage{url,hyperref,lineno,microtype,subcaption}
\usepackage[onehalfspacing]{setspace}

\def\keyFont{\fontsize{8}{11}\helveticabold }
\def\firstAuthorLast{Kora {et~al.}} 
\def\Authors{Youssef Kora\,$^{1,2,3,4*}$ and Christoph Simon\,$^{1,2,3}$}
\def\Address{$^{1}$Department of Physics and Astronomy, University of Calgary, Calgary, Alberta, Canada \\
$^{2}$Institute for Quantum Science and Technology, University of Calgary,  Calgary, Alberta, Canada  \\
$^{3}$Hotchkiss Brain Institute, University of Calgary,  Calgary, Calgary, Alberta, Canada  \\
$^{4}$Department of Physics, University of New Brunswick, Fredericton, New Brunswick, Canada }
\def\corrAuthor{Youssef Kora}

\def\corrEmail{ykora.ph@gmail.com}

\begin{document}

\onecolumn
\firstpage{1}

\title{Quantumness can enhance resilience to statistical noise in spin-network quantum reservoir computing.}


\author[\firstAuthorLast ]{\Authors} 
\address{} 
\correspondance{} 

\extraAuth{}

\maketitle

\begin{abstract}

Quantum reservoir computing offers a promising approach to the utilization of complex quantum dynamics in machine learning. Statistical noise inevitably arises in real settings of quantum reservoir computing (QRC) due to the practical necessity of taking a small to moderate number of measurements. We investigate the effect of statistical noise in spin-network QRC on the possible performance benefits conferred by quantumness. As our measures of quantumness, we employ both quantum entanglement, which we quantify by the partial transpose of the density matrix, and coherence, which we quantify as the sum of the absolute values of the off-diagonal elements of the density matrix. We find that reservoirs which enjoy a finite degree of quantum entanglement and coherence are more stable against the adverse effects of statistical noise on performance compared to their unentangled, incoherent counterparts. Our results thus indicate that the potential benefit reservoir computers may derive from quantumness depends on the number of measurements used for training and testing, and that statistical noise, albeit detrimental on the whole, may leave quantum reservoirs in a stronger position relative to less quantum ones. These findings not only emphasize the importance of incorporating realistic noise models, but also suggest that the search for computational regimes that benefit from quantumness may be aided rather than impeded by the practical constraints of implementation within existing machines.
\section{}

\tiny
 \keyFont{ \section{Keywords:} quantum reservoir computing, quantum AI, quantum entanglement, statistical noise, spin networks} 
\end{abstract}

\section{Introduction}

As we advance further into the information age, it is imperative that we should scale up our ability to interact with data of unprecedented complexity and volume. It is thus of increasingly critical importance to endow machines with the ability to understand, manipulate, and respond to sequential data inputs with exceeding efficiency. A primary challenge in this arena arises in the form of the so-called von Neumann bottleneck, in which the physical division of processing and memory units constrains processing speed \cite{von_Neumann_1993}. In stark contrast, biological systems manage, by integrating processing and memory units, dynamic, continuous information processing with vastly superior efficiency and remarkably low energy usage \cite{Eliasmith_2012,Stewart_2012}. \\ \indent
Reservoir computing (RC) is as a paradigm that promises to empower our machines in the face of these challenges \cite{jaeger_2004,maass_2002,verstraeten_2007}. At the heart of it is the reservoir, a dynamical system of very high dimensionality, which receives incoming streams of data and naturally generates transient internal states endowed with fading memory and non-linear processing capabilities. Such dynamical complexity lends itself well to machine learning tasks requiring a strong capacity for retaining prior inputs, such as speech recognition, stock market prediction, and autonomous motor control for robots \cite{Tanaka_2019}.  Early implementations of RC used randomly connected artificial neural networks or spiking neural networks \cite{Nicola_2017}, while physical realizations have emerged across various platforms, such as photonics \cite{Vandoorne_2014,Antonik_2017,Larger_2017,Sunada_2021,Garc_2023}, phonons \cite{Dion_2018,Meffan_2023}, magnons \cite{Papp_2021,Gartside_2022,K_rber_2023}, spintronics \cite{Torrejon_2017,Furuta_2018,Tsunegi_2018}, and neuromorphic nanomaterials \cite{Stieg_2011,Yaremkevich_2023}.
\\ \indent
\begin{figure}[h]
\centering
\includegraphics[width=0.5\textwidth]{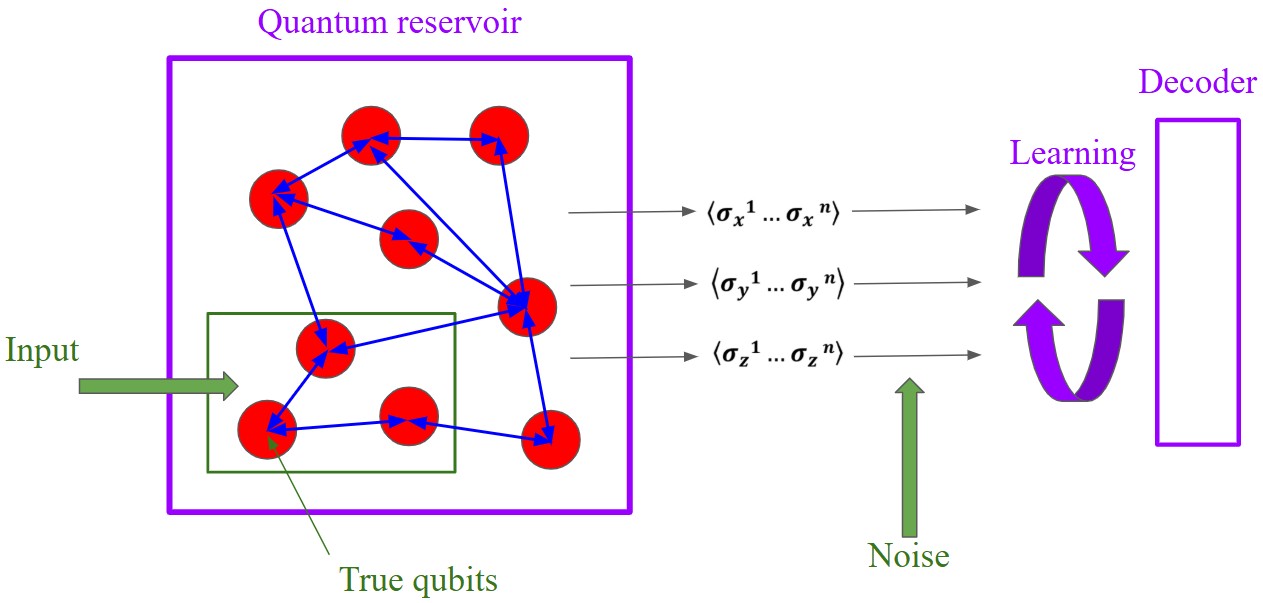}
    \caption{\textbf{Schematic representation of spin-network quantum reservoir computing.} The reservoir comprises $N$ qubits with random and unchanging connections therebetween. At each timestep, the input data is injected into a fraction of the qubits, true qubits. Following injection, a finite number of measurements is collected, which gives rise to statistical noise. The next stage is the learning process, which is undertaken only by the classical connections between the reservoir and the decoder.}.
\label{schematic}     
\end{figure}
Recent years have witnessed the rise of \textit{quantum} reservoir computing (QRC), which seeks to employ within the framework of RC the unique properties of quantum mechanics, such as quantum entanglement and coherence \cite{fujii_2017,fujii_2021}. One model of QRC utilizes networks of qubits \cite{fujii_2021,Luchnikov_2019,martinez_2020}, and another employs reservoirs comprising quantum-mechanical oscillators \cite{Govia_2021,Nokkala_2021,Dudas_2023}. Here, we focus on the qubit-based approach, in which changes to qubit states are driven by sequential inputs that propagate through the reservoir’s quantum dynamics \cite{Nakajima2019,martinez_2021}. Time-series data enters the system by injection into the so-called true nodes. The system, comprising both true nodes and hidden nodes, is allowed to evolve under the influence of the input. Finally, an output signal is read from all nodes, to be used in the learning stage. Crucially, training only takes place at the output layer, which simplifies the learning considerably. A schematic of this general QRC mechanism is shown in Fig. \ref{schematic}. \\ \indent
The chief obstacle in the face of quantum computing and quantum machine learning is the decoherence inherent to the operation of the available quantum devices. Significant efforts have been dedicated to correcting or mitigating the resulting errors. However, there have been indications that, in certain scenarios, this dissipative noise may be harnessed to provide a computational benefit in the context of QRC \cite{Fry2023, suzuki2022,Domingo2023,sannia_2024}. Another recent observation of a potential slight enhancement of the performance of weakly-interacting spin-network QRC was attributed to dissipation \cite{kora_2024}.\\ \indent
There has also been a rising interest in understanding the potential advantage conferred by the essential aspects of quantum mechanics — quantum entanglement and superposition — in QRC. Such an advantage stands in contrast with the rather more traditional advantage of the exponential scaling of the state space, which can be present in the absence of quantumness \cite{ehlers_2025}. Considerable research efforts have endeavored to investigate how quantumness might provide benefit to machine learning, in distributed learning over quantum networks \cite{gilboa_2023}, spin-network QRC \cite{gotting_2023}, and oscillator-based QRC \cite{motamedi_2023}. For example, Ref. \cite{kora_2024}, which sought to understand the physical circumstances conducing to an entanglement advantage in spin-network QRC, demonstrated that the presence of the entanglement advantage was, in the presence of dissipation, dependent on the frequency scale at which the input signal varies. This was interpreted as a consequence of the timescale introduced by dissipation, which determines whether quantum memory in the system can survive for long enough in the system for the input to manifest its temporal features, allowing the quantum reservoir computer to remember them. \\ \indent
Another important frontline in QRC is the ability to implement them on real machines. There are a number of candidate platforms which lend themselves well to the task, such as Rydberg atoms \cite{araiza_2022}, superconducting qubits \cite{suzuki_2022}, photonics \cite{Garc_2023}, and trapped ions \cite{haffner_2008}. Making contact with such implementations, however, faces a number of challenges, such as the necessity of contending with the measurement problem of quantum mechanics. Traditional QRC relies on projective measurements which destroy the quantum state, and thus require rewinding the system and input back in time every time a measurement is made. This, in combination with the necessity of repeating the process many times to compute quantum expectation values, leads to unfavorable time complexity and the need for an external memory. A number of approaches have been proposed to address this, such as those utilizing weak measurements \cite{Mujal_2023,franceschetto_2024}, feedback protocols \cite{kobayashi_2024}, an approach that combines the two \cite{monomi_2025}, and artificial memory restriction \cite{cindrak_2024}. However, it is generally the case that a real implementation will be rather severely constrained in the number of measurements that can be practically extracted from the machine. \\ \indent
In this work, we set out to explore the consequences of the statistical error introduced by the necessity of making a limited number of measurement in spin-network QRC, inspired by Ref. \cite{palacios_2024}. Given certain conditions in which quantumness can confer computational benefits \cite{kora_2024}, we are now interested in understanding how this error interacts with these benefits. We find that these potential benefits offered by quantumness (which we measure through entanglement and coherence) are qualitatively dependent on the strength of this statistical noise; while having an overall adverse effect on performance, statistical noise is less detrimental to reservoir computers with higher amounts of quantum entanglement and coherence than reservoirs which are unentangled and incoherent. Our results not only highlight the importance of accounting for statistical noise in QRC simulations, but also demonstrate that the necessity of collecting fewer measurements may actually lead QRC systems to derive a relative benefit from quantumness. Indeed, the presence of this statistical noise was found, under certain conditions, to turn a negative relationship between performance and quantumness into a positive one.\\ \indent
The effect of noise in the context of quantum machine learning has been explored in prior studies, such as  Ref. \cite{fangjun_2023}. In that work, they establish general upper bounds on the resolvable expressive capacity under sampling noise. Our study complements this by investigating a concrete spin‐network quantum reservoir (transverse‐field Ising), evaluating memory‐capacity performance, and demonstrating that increasing quantumness (entanglement and coherence) enhances resilience to measurement noise; we observe that under moderate levels of statistical noise, a negative trend between performance and quantumness can transform into a positive one; a ‘noise‐enabled’ effect. Thus does our work offer a distinct and complementary contribution, by explicitly linking quantum resource metrics with noise‐robust reservoir computing performance. \\ \indent
The remainder of this paper is organized as follows: in section \ref{meth} we describe our dynamical models and methodology. We present and discuss our results in section \ref{res}, and finally outline our conclusions in section \ref{conc}.

\section{Model and Methodology}\label{meth}
\subsection{Physical System}\label{sys}
\begin{figure}[h]
\centering
\includegraphics[width=0.7\textwidth]{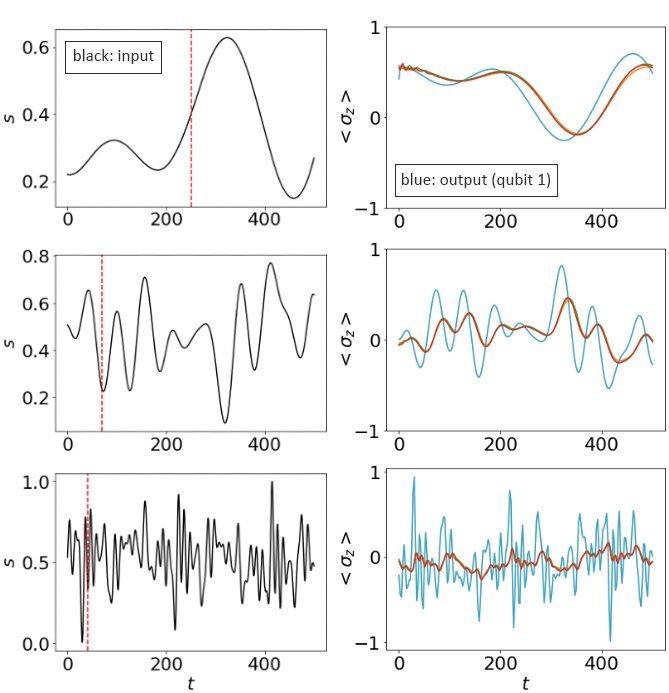}
    \caption{(left) Examples of input sequences at three different frequency scales $f=0.2$ (top), $f=1$ (middle), $f=5$ (bottom). The red vertical dashed line corresponds to the maximum time delay at which the reservoir is successful in remembering past input. (right) The corresponding outputs of the reservoir at an interaction strength of $J_s=1$, a transverse field of $h=2$, an injection period of $\Delta=2.5$, and a dissipation strength of $\Gamma=0.01$. Blue corresponds to the input qubit. Reproduced from Ref. \cite{kora_2024}(© 2024 American Physical Society), with permission.}
\label{input}     
\end{figure}

Our physical model of the quantum reservoir follows our previous formulation in Ref. \cite{kora_2024}. Figures \ref{schematic} and \ref{input} are reproduced from that work for clarity. It is a network of  $N=4$ qubits obeying a generalized transverse-field Ising model \cite{stinchcombe_1973,pfeuty_1971} with the Hamiltonian
\begin{eqnarray}\label{ham}
\hat H = \sum_{i>j=1}^{N} J_{ij} \hat\sigma_{i}^{x} \hat\sigma_{j}^{x} + h \sum_{i=1}^{N} \hat\sigma_{i}^{z},
\end{eqnarray}
where $\hat \sigma_i^a$ ($a=x,y,z$) are the Pauli operators, $h$ is the transverse magnetic field, and $J_{ij}$ are randomly generated network connectivities sampled from a uniform distribution in the interval $[-J_s/2,J_s/2]$.
Ours is an open quantum system experiencing Markovian dynamics and obeying the Lindblad master equation 
\begin{align}\label{mastereq}
\frac{d \hat\rho}{dt} = \mathcal{\hat L} \hat\rho = -i[\hat H, \hat\rho] + \Gamma \sum_{i=1}^{N} \left( \hat L_i \hat \rho \hat L_i^\dagger - \frac{1}{2} \{\hat L_i^\dagger \hat L_i, \hat \rho\} \right),
\end{align}
$\Gamma$ being the dissipation rate in this high-temperature decoherence channel \cite{breuer_2002}, and $\{\hat L_i\}$ are identifiable as the raising and lowering operators of the system.
The unitary system experiences a dynamical phase transition between an ergodic phase and a many-body-localized phase determined by the ratio $h/J_s$ \cite{martinez_2021}. 
\subsection{Input, Training, and Statistical Noise}\label{inp}
As in Ref. \cite{kora_2024}, we utilized input signals $s_k$ with a frequency scale $f$, defined as the sum of 20 signals of equal amplitude and frequencies $\left\{f_i\right\}$ chosen with equal linear spacing in the interval $\left[ f/5000, f/50 \right]$:
\begin{align}\label{sk}
s_k = \sum_i^{20} \sin\left(2\pi f_i  t_k + 2 \pi \zeta\right),
\end{align}
where $t_k$ is the time after $k$ time steps, and $\zeta$ is sampled from the uniform distribution in the interval $\left[0,1\right]$. Input signals are normalized so as to lie between 0 and 1. The left panels of Fig. \ref{input} show examples of these inputs, and the right panels show the corresponding outputs of the quantum reservoir at a representative choice of parameters.
This form of the input signal allows us to tune its frequency scale, which is conducive to our goal of explicitly addressing the interplay between the input frequency and the dissipation time scale discussed in Ref. \cite{kora_2024}. \\ \indent 
The input signals are introduced into the quantum reservoir by means of reinitializing the state of one qubit that we call qubit 1, a commonly-used form of input injection \cite{mujal_2021}: we trace out the qubit in question and prepare it in the input-defined state $|\psi_{s_k}\rangle = \sqrt{1 - s_k} |0\rangle + \sqrt{s_k} |1\rangle$, thereby transforming the quantum state of the system $\rho$ into the product state
\begin{align}\label{fourier2}
\hat \rho \rightarrow |\psi_{s_k}\rangle \langle\psi_{s_k}| \otimes \text{Tr}_{1}[\hat \rho].
\end{align}
This injection occurs ever $\Delta t =L\delta t$, where $L$ is the number of time steps between successive input injections, chosen so as to lie in the favorable range reported in Ref. \cite{martinez_2020} in the interest of rich dynamical behavior (here taken to be $\Delta t$ = 2.5 and $L$ = 100).   \\ \indent
Our reservoirs are time-multiplexed; the outputs are extracted at time intervals of $\Delta t/V$, giving rise to $NV$ virtual nodes (here $V=10$). The readout  signals are here restricted to $\langle \sigma_z\rangle$, and are trained in a linear regression to produce the desired function of the input, which in this work is the fundamental linear memory task \cite{carroll_2022} $\bar{y}_k = s_{k-\tau}$ ($\bar{y}_k$ being the target).  \\ \indent
After training the system on multiple sequences, we test it using a different sequence of the same frequency scale $f$. We employ Tikhonov regularization in which a regularization parameter is chosen to maximize performance. The performance of the reservoir in a memory task with time delay $\tau$ is called the memory capacity, which we compute as
\begin{align}\label{memcap}
C^{\tau}_{STM} = \frac{\text{cov}^2(y, \bar{y}^\tau)}{\sigma_{y}^2 \sigma_{\bar{y}^\tau}^2},
\end{align}
where  $y$ is the output signal of the reservoir and $\bar{y}^\tau$ is the target function at a time delay $\tau$. \\ \indent
The output signals of the reservoir, being a product of our simulations rather than real machines, are subjected to a level of Gaussian statistical noise $\sigma$ to simulate the necessity of taking a finite number of measurements when using a real machine as a quantum reservoir computer. The value of $\sigma$, whose effect on the physics is a primary subject of our investigation, corresponds to the number of measurements that would be made in a real setting. As in Ref. \cite{palacios_2024}, the Gaussian standard deviation is taken to be inversely proportional to the square root of the number of measurements , i.e.,  $ \propto 1/\sqrt{N}$. Thus, for example, $ \sigma =0.001$ would represent the physical situation of taking a million measurements. \\ \indent
The quantum reservoir computer is characterized by the fading memory property \cite{dambre_2012}, which guarantees that the memory capacity vanish at large $\tau$. Thus, the total memory capacity of the reservoir computer may be defined as
\begin{align}\label{totmemcap}
C_{\text{STM}} = \sum_{\tau=0}^{\infty} C_{\text{STM}}^{\tau}.
\end{align}
Memory capacities are typically computed for inputs which are independent and identically drawn from a probability distribution (i.i.d.), in which case the total memory capacity is bounded by the number of linearly independent state variables of the system \cite{dambre_2012}. Our input sequences are not i.i.d., but contain periodic structures (as do most real-world sequences), and are thus not subject to that upper bound. On that account, a numerical comparison between the total memory capacity of such periodic sequences and standard i.i.d. benchmarks must be carried out with care. For instance, in Ref. \cite{kora_2024}, comparisons between capacities of different signals are undertaken by normalizing by the frequency scale.
\subsection{Entanglement and Coherence}\label{pca}
\begin{figure}[h]
\centering
\includegraphics[width=0.5\textwidth]{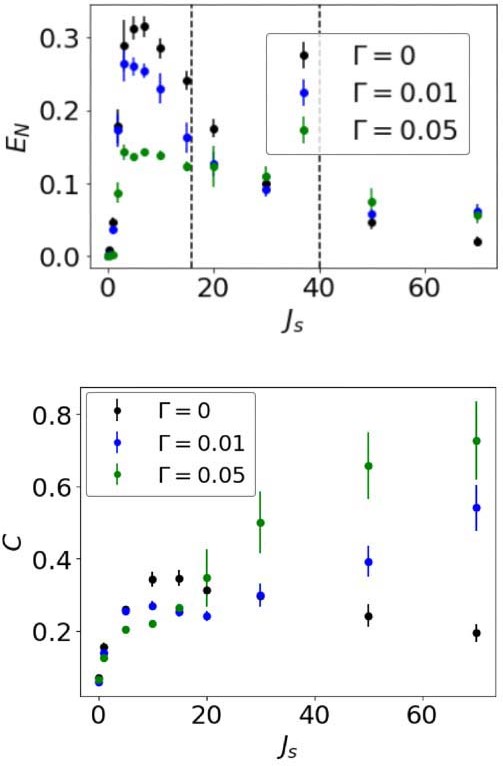}
    \caption{(top) Entanglement vs. interaction strength for the reservoir at a transverse field of $h=2$, and an injection period of $\Delta t=2.5$.  The input of frequency of the signal is fixed at $f=0.2$ while the dissipation strength $\Gamma$ is varied. (bottom) Coherence vs. interaction strength for the reservoir at the same physical conditions, at various dissipation strengths  and constant frequency. }
\label{entco}     
\end{figure}
We produce reservoir computers of varying entanglement and coherence by varying the interaction strength $J_s$, and measure the amount of quantum entanglement therein by means of the logarithmic negativity \cite{vidal_2002,plenio_2005}
\begin{equation}\label{logent}
E_N(\rho) = \log_2 \|\rho^{\Gamma_A}\|_1,
\end{equation}
where $\| \cdot \|_1$ is the trace norm, $\rho^{\Gamma_A}$ is the partial transpose with respect to subsystem A, and all possibe bipartitions of the system are averaged over. \\ \indent
We have chosen to employ logarithmic negativity as our chief entanglement measure because it is a computable entanglement monotone that is well suited to mixed states and open quantum systems, requiring only the partial transpose of the density matrix. It is thus a widely used and numerically tractable indicator of bipartite quantum correlations in dynamical systems such as the one considered here. For an outlook towards more nuanced alternatives, see Sec. \ref{conc}.
\\ \indent 
Because of our mechanism of input injection, quantum entanglement abruptly plummets each time the input qubit (qubit 1) is reinitialized to reflect the input signal. This drop is naturally strongest for the entanglement corresponding to the bipartition isolating qubit 1. We show an example of this behaviour in Fig. \ref{entco} (top panel), where 4 different bipartitions are shown, each isolating one of the 4 qubits.  \\ \indent
As our secondary measure of quantumness, we quantify the coherence of a quantum state $\rho$ by means of the $l_1$-{\em norm of coherence} \cite{streltsov_2017, baumgratz_2014}, which is given by
\begin{equation}\label{coh}
C = \sum_{i \neq j} |\rho_{ij}|,
\end{equation}
and is normalized by its maximum value of $2^N -1$.
\section{Results and Discussion}\label{res}
We start by reviewing the behavior of the logarithmic negativity measure of entanglement. In Fig. \ref{entco} (middle panel), we present it as a function the interaction strength for a variety of dissipation strengths, at a transverse field of $h=2$, and an injection period of $\Delta t=2.5$. The system undergoes a dynamical phase transition between an ergodic and many-body localized phase  \cite{martinez_2021}, marked by the region between the black, dashed lines. Entanglement peaks to the left of the transition, i.e., in the ergodic regime, and goes to zero at both extremes. It is also worth remarking that the peak is suppressed as $\Gamma$ rises, but the frequency of the input signal has little to no effect. \\ \indent
To contrast, we examine the behavior of the coherence of our reservoirs in Fig. \ref{entco} (bottom panel). These results suggest that coherence exhibits a peak as a function of interaction strength only in the case of the isolated reservoir, whereas open systems gain coherence in the many-body localized phase. This behavior is consistent with Lindbladian many-body localization \cite{hamazaki_2022}, in which interactions may prevent many-body decoherence. In such a regime, long-range entanglement is degraded by the environment, while coherence remains strong. On the other hand, coherence appears to be insensitive to the frequency of the input, as is entanglement.\\ \indent
Fig. \ref{entco} illustrates that quantumness measures such as entanglement and coherence are determined by our choice of the parameters that govern the system's dynamics. The presence of such quantumness measures can correspond to better performance, depending, as shown in Ref. \cite{kora_2024}, on the interplay between the dissipation time scale and the input frequency scale. In Ref. \cite{kora_2024}, such a phenomenon was referred to as an entanglement advantage. We henceforth use that term (entanglement advantage) to refer to a situation where entangled reservoirs exhibit better performance at the task at hand, compared to the same reservoir with low entanglement. \\ \indent
\begin{figure}[t]
\centering
\includegraphics[width=0.8\textwidth]{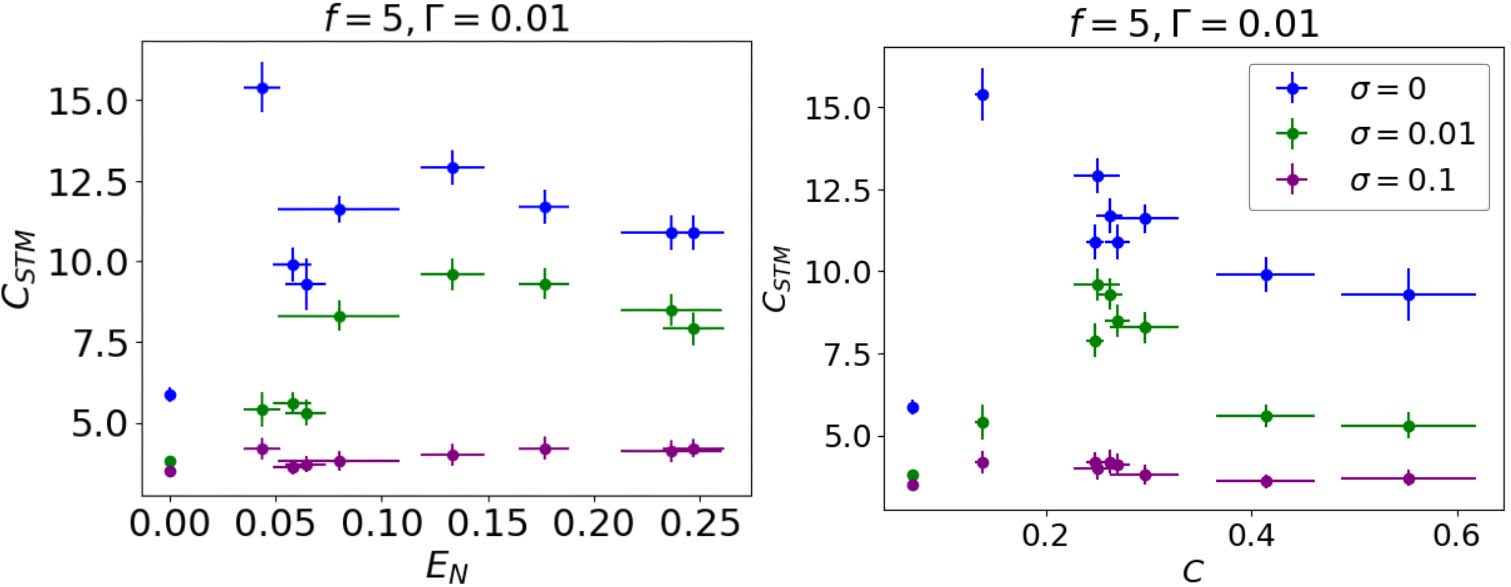}
    \caption{The high-frequency case: The effect of the statistical noise level $\sigma$ on the total memory capacity as a function of entanglement (left) and coherence (right). The dissipation strength is $\Gamma=0.01$, and the input of frequency of the signal is $f=5$. The transverse field is fixed at $h=2$, and the injection period at $\Delta t=2.5$.}
\label{f5g1}     
\end{figure}
\begin{figure}[t]
\centering
\includegraphics[width=0.8\textwidth]{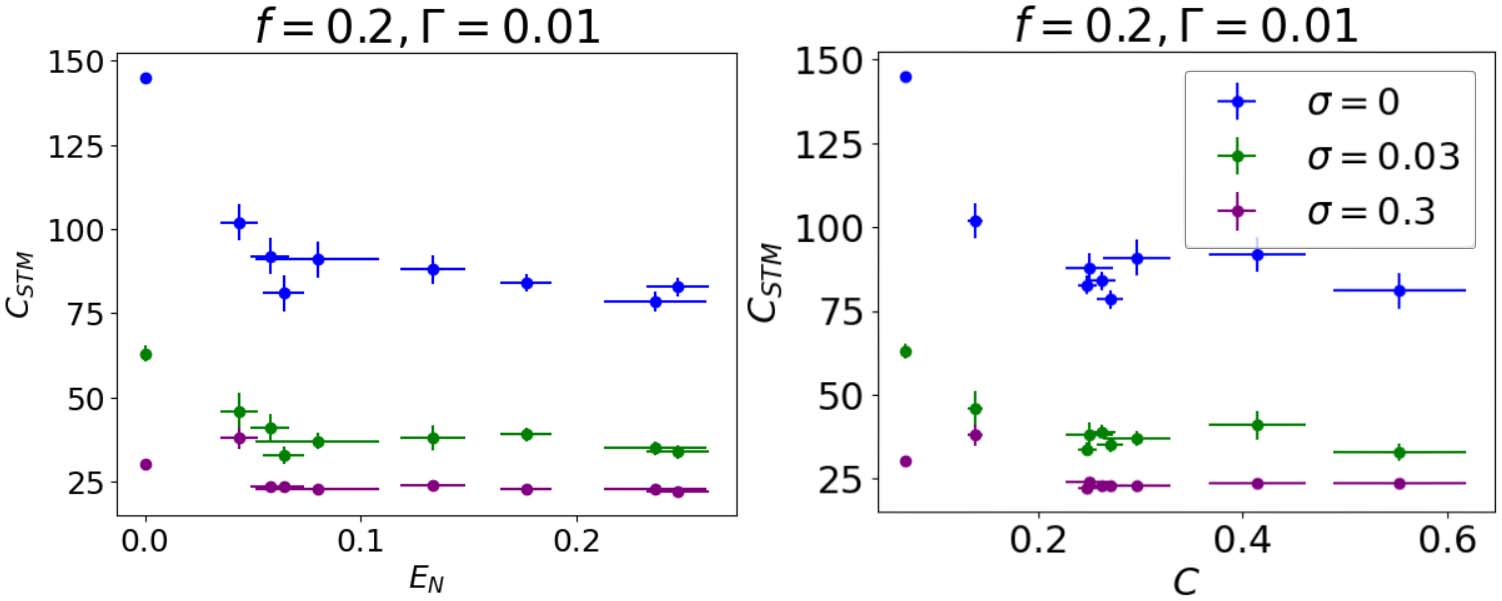}
    \caption{The low-frequency case: The effect of the statistical noise level $\sigma$ on the total memory capacity as a function of entanglement (left) and coherence (right). The dissipation strength is $\Gamma=0.01$, and the input of frequency of the signal is $f=0.2$. The transverse field is fixed at $h=2$, and the injection period at $\Delta t=2.5$.}
\label{f0p2g1}     
\end{figure}
\begin{figure}[t]
\centering
\includegraphics[width=0.8\textwidth]{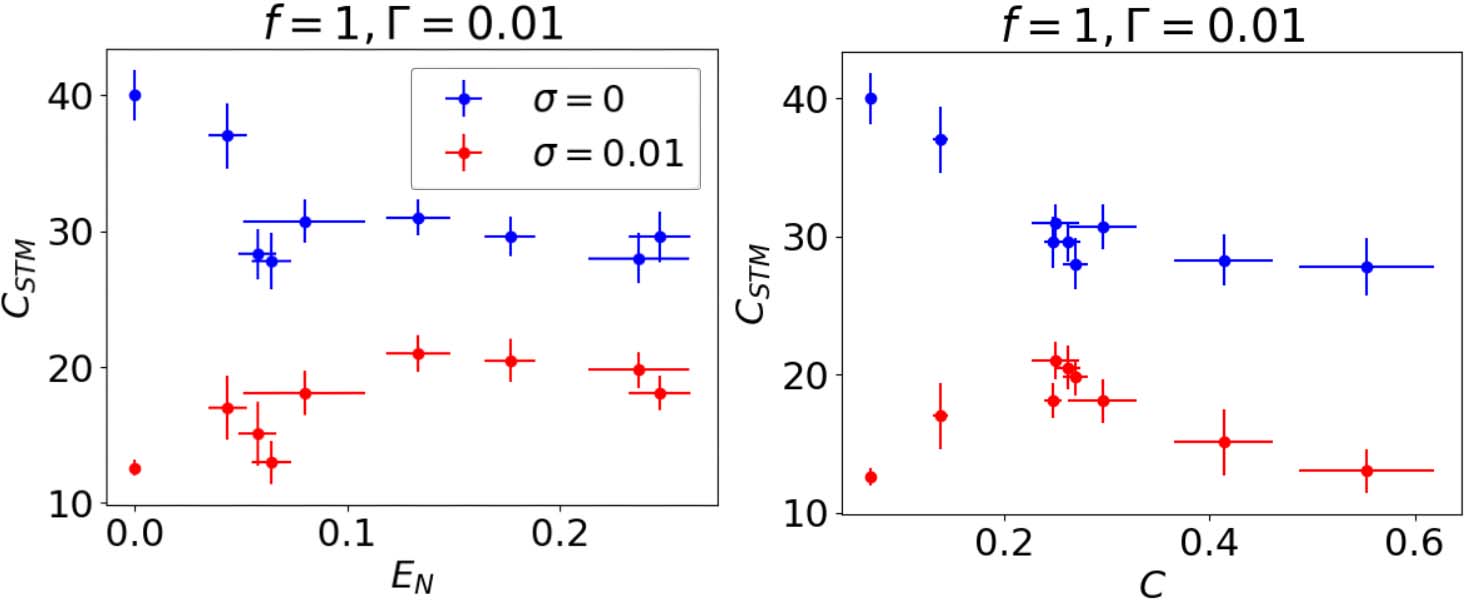}
    \caption{The intermediate-frequency case: The effect of the statistical noise level $\sigma$ on the total memory capacity as a function of entanglement (left) and coherence (right). The dissipation strength is $\Gamma=0.01$, and the input of frequency of the signal is $f=1$. The transverse field is fixed at $h=2$, and the injection period at $\Delta t=2.5$.}
\label{f1g1}     
\end{figure}
   
\begin{figure}[t]
\centering
\includegraphics[width=0.8\textwidth]{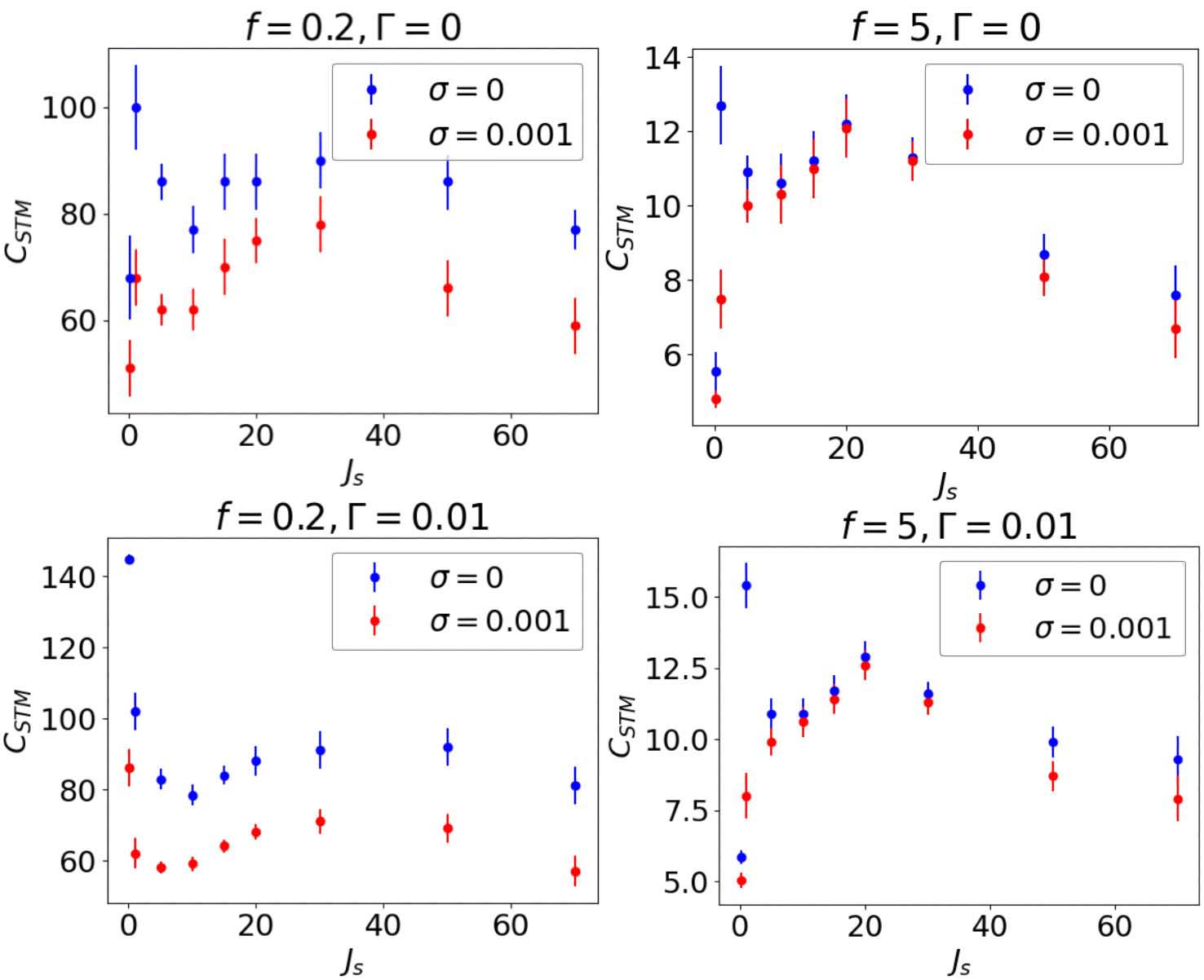}
    \caption{The effect of the statistical noise level $\sigma$ on the total memory capacity  as a function of interaction strength, at two different dissipation strengths $\Gamma$ and two input frequencies $f$. The transverse field is fixed at $h=2$, and the injection period at $\Delta t=2.5$.}
\label{CvJ}     
\end{figure}
We now turn to the effect of Gaussian noise resulting from a finite number of measurements at the output layer. We look at conditions corresponding to both an entanglement advantage (high input frequency in the presence of finite dissipation) and an entanglement disadvantage (low input frequency in the presence of finite dissipation), as detailed in Ref. \cite{kora_2024}. As noted in Sec. \ref{meth}B, our sequences are not i.i.d., and memory capacities are much higher at lower frequencies due to the slower variation of the input signal.  \\ \indent
In Fig. \ref{f5g1}, we intensify the effect of statistical noise (which amounts to making fewer measurements) in the high-frequency regime that decidedly performs better in the presence of quantum entanglement and  coherence, as can be seen by comparing the performance of the leftmost points compared to the rest. This high frequency is chosen to be $f=5$ at a dissipation strength of $\Gamma=0.01$. We present the total memory capacity for a wide range of statistical noise levels. Our most intense statistical noise level is $\sigma=0.1$. The memory capacity is invariably reduced by increasing the statistical noise, but the amount of performance loss is dependent on the amount of entanglement and coherence in the reservoir. \\ \indent
This observation raises an important question: what happens to the entanglement advantage as the rising statistical noise level causes the performance to drop? Our results in the left panel of Fig. \ref{f5g1} indicate that the entanglement advantage very much persists, notwithstanding the overall reduction in memory capacity. We also found it interesting to look at how performance behaves as a function of coherence rather than entanglement. These results are shown in the right panel of the same figure, and they tell a similar story, indicating that both entanglement and coherence can function as measures of quantumness whose presence may correspond to a computational advantage on reservoir computers. Specifically, there is a peak for both entanglement and coherence, but the peak is more pronounced as a function of coherence; this may be attributable to the continued growing of coherence in the localized regime. There seems to be an {\em excessive} amount of quantumness, as measured by either entanglement or coherence, that does not correspond to much benefit. But since coherence continues to grow in the many-body localized phase while entanglement dies, as may be seen in Fig. \ref{entco}, the performance peak is more defined with coherence on the x-axis. \\ \indent
A most interesting question presents itself at this point: what happens in the {\em low-frequency} (and finite dissipation) regime which does not appear to, in the absence of statistical noise, allow the system to derive much benefit from quantumness? In Fig. \ref{f0p2g1}, we present our calculations of the total memory capacity for an input frequency of $f=0.2$ and a dissipation strength of $\Gamma=0.01$. While ramping up the statistical noise unsurprisingly lowers performance throughout, it is clear from the figure that the high-entanglement, high-coherence reservoirs are more durable against the adverse effect of statistical noise. The points with low entanglement and coherence, on the other hand, are brought down the most as a function of statistical noise. Indeed, the negative trend in the case without statistical errors does not persist as the number of measurements becomes of the order of 100, a peak having emerged at moderate levels of quantumness. \\ \indent
The behavior above evinces the possibility of turning an unfavorable trend into an unambiguously {\em positive} one. To further investigate this effect, we turn to the case of the intermediate input frequency of $f=1$, whose results we present in Fig. \ref{f1g1}. Remarkably, the quantum indifference (or slight disadvantage) observable without accounting for statistical noise does indeed turn into a distinctly favorable trend. At this level of statistical noise, reservoirs with finite quantum entanglement and coherence do perform better than the unentangled, incoherent reservoir. However, performance saturation appears more distinctly in the case of quantum entanglement on the x axis, whereas there seems to be rather a decline in the case of coherence, similar to the high-frequency case. Thus, Fig. \ref{f1g1} presents a phenomenon where statistical noise causes a decrease in overall performance but the emergence of a relative advantage for reservoirs with more quantumness. \\ \indent
We make one last point by looking at memory capacity as a function of interaction strength in Fig. \ref{CvJ}, both in the presence and the absence of noise ($\sigma=0.001$, which corresponds to a million measurements). As established, statistical noise invariably reduces performance, albeit to greatly varying extents. The low-frequency regime appears to generally sustain a greater performance reduction, both in the presence and absence of dissipation. The high-frequency regime, on the other hand, exhibits a performance loss only at the spike at low but non-zero interaction strength. This spike is present only in the absence of statistical noise, and is more pronounced in the presence of dissipation, as reported in Ref. \cite{kora_2024}. By considering only the case without statistical noise, one may be tempted to conclude that dissipation might offer a computational advantage in that regime. However, our results here, which show that such a performance improvement does not hold in the presence of statistical noise, corroborate the argument made in Ref. \cite{palacios_2024} respecting the necessity of accounting for statistical noise resulting from a finite number of measurements.
\section{Conclusions and Outlook}\label{conc}
Our central objective in this work was to investigate the behavior of our spin-network QRC system within the practical constraints of having to take a limited number of measurements. We studied the effect of the resulting statistical noise on the performance of our system in linear memory tasks, and its implications with respect to the frequency of the input signal, the presence of dissipation in the system, and quantumness as measured by quantum entanglement and coherence. Our chief conclusion is that, while the presence of statistical noise degrades performance on the whole, reservoirs that exhibit quantum entanglement and coherence are more resistant to this degradation that inevitably arises in implementations on present-day machines. \\ \indent
A more realistic model of statistical noise also provides protection against possible spurious observations of performance enhancements due to, for instance, dissipation at low interaction strength. This emphasizes the importance of accounting for the reality of a finite number of measurements when employing real machines for QRC. \\ \indent
In general, performance in general declines when the system is afflicted with the statistical noise resulting from a finite number of measurements. However, substantially quantum reservoirs are found to be more resistant to the effect of statistical noise than their non-quantum counterparts. As a result, it appears that, under the constraint of having to take a small to moderate number of measurements, we observe an entanglement advantage in certain regimes in which none was observed in the absence of statistical noise.
\\ \indent
It is worth clarifying our use of the term "entanglement advantage", which in this study refers to the performance comparison between highly entangled states and low-entanglement states within quantum reservoirs, rather than an advantage over classical algorithms. It describes a phenomenon in which there is an observed improvement in QRC performance when there is non-zero entanglement. It does not, however, refer to a monotonic relationship where entanglement is the sole determiner of performance. For example, it is clear that stronger coupling between qubits can by itself influence reservoir memory by altering dynamical timescales; however, the observed trends across different interaction and dissipation regimes do not suggest that coupling is solely responsible either. For instance, it is clear from the right panels of Fig. \ref{CvJ} that high interaction strength leads to a decline in total memory capacity. Fig. \ref{f5g1} also shows that performance is not monotonically related to coherence and entanglement either, but that entangled and coherent reservoirs can be more resilient to statistical noise. The relative extents to which coupling and quantumness respectively correlate with performance is beyond the scope of the present study, but it is an interesting question for future investigation \\ \indent
It is also worth taking a moment to emphasize the distinction between our demonstrated quantum enhancement in noise tolerance on one hand, and the quantum advantage on the other. The latter refers to a computational advantage derived from quantum effects which cannot be replicated in a classical system, and its thorough demonstration is an ongoing quest. While our results here provide important considerations for that quest, they must be interpreted as evidence of entanglement-enhanced statistical noise resilience in quantum reservoir computers rather than of absolute quantum computational supremacy over classical reservoir computers.
\\ \indent
We must stress again that our use of terms like “entanglement advantage” refers to regimes where entangled reservoirs enjoy a computational advantage relative to their non-entangled version, the latter still being a quantum system albeit in its classical limit. Such a quantum reservoir in its classical limit is still afflicted with the inevitability of statistical noise, and so incorporating it into the analysis is important for investigating the degree to which more entangled reservoirs may perform better than those with little to no quantum entanglement. 
\\ \indent
Our observed resilience to statistical noise is evocative of a similar phenomenon in continuous-variable QRC, where the quantum resource of squeezing was found to enhance the robustness of the reservoir to readout noise, giving rise to performance enhancement in linear and non-linear memory tasks \cite{garcia_2024}. 
\\ \indent
It is important to note that our model of statistical errors here does not lead to an uncertainty in the quantum state, for which we have an exact description by means of solving eq. \ref{mastereq}. In a true experiment, however, the quantum state must be inferred from the measurements, and so the level of statistical noise would affect one's knowledge of the quantum state, and hence of the quantumness. In situations where quantumness cannot be measured to the desired precision, it is unclear whether its benefits may be meaningfully evaluated. The implications of statistical errors on the ability to accurately measure quantumness is a topic for future investigation. 
\\ \indent
It is also worth reflecting on our reliance on primarily entanglement (and secondarily coherence) to quantify quantumness within our reservoirs. It may be of great interest to examine how this reliance may be affected by the existence of classically simulable operations with a high degree of entanglement, such as stabilizer states and their associated Clifford circuits \cite{gu_2025}. A more nuanced measure of quantumness in light of this is the so-called magic, or nonstabilizerness, which may have implications for the possibility of reproducing the observed performance classically \cite{maronese_2025}. A future study could employ magic quantifiers, such as the stabilizer 2-Rényi entropy \cite{leone_2022}, to systematically analyze its presence and impact within our spin-network QRC scheme and how non-stabilizer operations might correlate with enhanced resilience to statistical noise.
\\ \indent In short, our results suggest that, within the limitations discussed above, the presence of statistical noise, while on the whole detrimental to performance, may nevertheless elevate the system from a regime that does not benefit from quantumness, as measured by entanglement and coherence, to a regime that does. This suggests that constraints imposed by real implementations of QRC may, rather surprisingly, serve rather than obstruct the search for QRC systems in which the presence of quantumness corresponds to higher performance.

\section*{Author Contributions}

YK: conceptualization, methodology, investigation, computation, code development, manuscript writing; CS:  conceptualization, supervision, validation, manuscript review and editing, funding acquisition.  All authors have read and agreed to the final version of the manuscript.

\section*{Funding}
This work was supported by the National Research Council through its Applied Quantum Computing Challenge Program, the Natural Sciences and Engineering Research Council (NSERC) of Canada through its NSERC Discovery Grant Program, the Alberta Major Innovation Fund, and Quantum City. 

\section*{Acknowledgments}
We would like to thank Khabat Heshami and Hadi Zadeh-Haghighi for the useful discussions and feedback.

\section*{Data Availability Statement}
The datasets generated and analyzed during the current study are available from the corresponding author upon reasonable request. 

\bibliographystyle{Frontiers-Harvard} 
\bibliography{biblio}

@article{haffner_2008,
title = {Quantum computing with trapped ions},
journal = {Physics Reports},
volume = {469},
number = {4},
pages = {155-203},
year = {2008},
issn = {0370-1573},
doi = {https://doi.org/10.1016/j.physrep.2008.09.003},
url = {https://www.sciencedirect.com/science/article/pii/S0370157308003463},
author = {H. H\"affner and C.F. Roos and R. Blatt},
}

@article{Suzuki2022,
  title = {Natural quantum reservoir computing for temporal information processing},
  volume = {12},
  ISSN = {2045-2322},
  url = {http://dx.doi.org/10.1038/s41598-022-05061-w},
  DOI = {10.1038/s41598-022-05061-w},
  number = {1},
  journal = {Scientific Reports},
  publisher = {Springer Science and Business Media LLC},
  author = {Suzuki,  Yudai and Gao,  Qi and Pradel,  Ken C. and Yasuoka,  Kenji and Yamamoto,  Naoki},
  year = {2022},
  month = jan 
}

@article{Fry2023,
  title = {Optimizing quantum noise-induced reservoir computing for nonlinear and chaotic time series prediction},
  volume = {13},
  ISSN = {2045-2322},
  url = {http://dx.doi.org/10.1038/s41598-023-45015-4},
  DOI = {10.1038/s41598-023-45015-4},
  number = {1},
  journal = {Scientific Reports},
  publisher = {Springer Science and Business Media LLC},
  author = {Fry,  Daniel and Deshmukh,  Amol and Chen,  Samuel Yen-Chi and Rastunkov,  Vladimir and Markov,  Vanio},
  year = {2023},
  month = nov 
}

@article{Domingo2023,
  title = {Taking advantage of noise in quantum reservoir computing},
  volume = {13},
  ISSN = {2045-2322},
  url = {http://dx.doi.org/10.1038/s41598-023-35461-5},
  DOI = {10.1038/s41598-023-35461-5},
  number = {1},
  journal = {Scientific Reports},
  publisher = {Springer Science and Business Media LLC},
  author = {Domingo,  L. and Carlo,  G. and Borondo,  F.},
  year = {2023},
  month = may 
}

@article{Yaremkevich_2023, 
    title={On-chip phonon-magnon reservoir for neuromorphic computing}, 
    volume={14}, ISSN={2041-1723}, url={http://dx.doi.org/10.1038/s41467-023-43891-y}, DOI={10.1038/s41467-023-43891-y}, number={1}, journal={Nature Communications}, publisher={Springer Science and Business Media LLC}, author={Yaremkevich, Dmytro D. and Scherbakov, Alexey V. and De Clerk, Luke and Kukhtaruk, Serhii M. and Nadzeyka, Achim and Campion, Richard and Rushforth, Andrew W. and Savel’ev, Sergey and Balanov, Alexander G. and Bayer, Manfred}, year={2023}, month=dec }

@article{Vandoorne_2014, title={Experimental demonstration of reservoir computing on a silicon photonics chip}, volume={5}, ISSN={2041-1723}, url={http://dx.doi.org/10.1038/ncomms4541}, DOI={10.1038/ncomms4541}, number={1}, journal={Nature Communications}, publisher={Springer Science and Business Media LLC}, author={Vandoorne, Kristof and Mechet, Pauline and Van Vaerenbergh, Thomas and Fiers, Martin and Morthier, Geert and Verstraeten, David and Schrauwen, Benjamin and Dambre, Joni and Bienstman, Peter}, year={2014}, month=mar }

@article{Antonik_2017, title={Brain-Inspired Photonic Signal Processor for Generating Periodic Patterns and Emulating Chaotic Systems}, volume={7}, ISSN={2331-7019}, url={http://dx.doi.org/10.1103/PhysRevApplied.7.054014}, DOI={10.1103/physrevapplied.7.054014}, number={5}, journal={Physical Review Applied}, publisher={American Physical Society (APS)}, author={Antonik, Piotr and Haelterman, Marc and Massar, Serge}, year={2017}, month=may }

@article{Larger_2017, title={High-Speed Photonic Reservoir Computing Using a Time-Delay-Based Architecture: Million Words per Second Classification}, volume={7}, ISSN={2160-3308}, url={http://dx.doi.org/10.1103/PhysRevX.7.011015}, DOI={10.1103/physrevx.7.011015}, number={1}, journal={Physical Review X}, publisher={American Physical Society (APS)}, author={Larger, Laurent and Baylón-Fuentes, Antonio and Martinenghi, Romain and Udaltsov, Vladimir S. and Chembo, Yanne K. and Jacquot, Maxime}, year={2017}, month=feb }

@article{Sunada_2021, title={Photonic neural field on a silicon chip: large-scale, high-speed neuro-inspired computing and sensing}, volume={8}, ISSN={2334-2536}, url={http://dx.doi.org/10.1364/OPTICA.434918}, DOI={10.1364/optica.434918}, number={11}, journal={Optica}, publisher={Optica Publishing Group}, author={Sunada, Satoshi and Uchida, Atsushi}, year={2021}, month=nov, pages={1388} }

@article{Dion_2018, title={Reservoir computing with a single delay-coupled non-linear mechanical oscillator}, volume={124}, ISSN={1089-7550}, url={http://dx.doi.org/10.1063/1.5038038}, DOI={10.1063/1.5038038}, number={15}, journal={Journal of Applied Physics}, publisher={AIP Publishing}, author={Dion, Guillaume and Mejaouri, Salim and Sylvestre, Julien}, year={2018}, month=oct }

@article{Meffan_2023, title={Non-linear processing with a surface acoustic wave reservoir computer}, volume={29}, ISSN={1432-1858}, url={http://dx.doi.org/10.1007/s00542-023-05463-4}, DOI={10.1007/s00542-023-05463-4}, number={8}, journal={Microsystem Technologies}, publisher={Springer Science and Business Media LLC}, author={Meffan, Claude and Ijima, Taiki and Banerjee, Amit and Hirotani, Jun and Tsuchiya, Toshiyuki}, year={2023}, month=may, pages={1197–1206} }

@article{Papp_2021, title={Nanoscale neural network using non-linear spin-wave interference}, volume={12}, ISSN={2041-1723}, url={http://dx.doi.org/10.1038/s41467-021-26711-z}, DOI={10.1038/s41467-021-26711-z}, number={1}, journal={Nature Communications}, publisher={Springer Science and Business Media LLC}, author={Papp, Adam and Porod, Wolfgang and Csaba, Gyorgy}, year={2021}, month=nov }

@article{Gartside_2022, title={Reconfigurable training and reservoir computing in an artificial spin-vortex ice via spin-wave fingerprinting}, volume={17}, ISSN={1748-3395}, url={http://dx.doi.org/10.1038/s41565-022-01091-7}, DOI={10.1038/s41565-022-01091-7}, number={5}, journal={Nature Nanotechnology}, publisher={Springer Science and Business Media LLC}, author={Gartside, Jack C. and Stenning, Kilian D. and Vanstone, Alex and Holder, Holly H. and Arroo, Daan M. and Dion, Troy and Caravelli, Francesco and Kurebayashi, Hidekazu and Branford, Will R.}, year={2022}, month=may, pages={460–469} }

@article{K_rber_2023, title={Pattern recognition in reciprocal space with a magnon-scattering reservoir}, volume={14}, ISSN={2041-1723}, url={http://dx.doi.org/10.1038/s41467-023-39452-y}, DOI={10.1038/s41467-023-39452-y}, number={1}, journal={Nature Communications}, publisher={Springer Science and Business Media LLC}, author={Körber, Lukas and Heins, Christopher and Hula, Tobias and Kim, Joo-Von and Thlang, Sonia and Schultheiss, Helmut and Fassbender, Jürgen and Schultheiss, Katrin}, year={2023}, month=jul }

@article{Torrejon_2017, title={Neuromorphic computing with nanoscale spintronic oscillators}, volume={547}, ISSN={1476-4687}, url={http://dx.doi.org/10.1038/nature23011}, DOI={10.1038/nature23011}, number={7664}, journal={Nature}, publisher={Springer Science and Business Media LLC}, author={Torrejon, Jacob and Riou, Mathieu and Araujo, Flavio Abreu and Tsunegi, Sumito and Khalsa, Guru and Querlioz, Damien and Bortolotti, Paolo and Cros, Vincent and Yakushiji, Kay and Fukushima, Akio and Kubota, Hitoshi and Yuasa, Shinji and Stiles, Mark D. and Grollier, Julie}, year={2017}, month=jul, pages={428–431} }

@article{Furuta_2018, title={Macromagnetic Simulation for Reservoir Computing Utilizing Spin Dynamics in Magnetic Tunnel Junctions}, volume={10}, ISSN={2331-7019}, url={http://dx.doi.org/10.1103/PhysRevApplied.10.034063}, DOI={10.1103/physrevapplied.10.034063}, number={3}, journal={Physical Review Applied}, publisher={American Physical Society (APS)}, author={Furuta, Taishi and Fujii, Keisuke and Nakajima, Kohei and Tsunegi, Sumito and Kubota, Hitoshi and Suzuki, Yoshishige and Miwa, Shinji}, year={2018}, month=sep }

@article{Tsunegi_2018, title={Evaluation of memory capacity of spin torque oscillator for recurrent neural networks}, volume={57}, ISSN={1347-4065}, url={http://dx.doi.org/10.7567/JJAP.57.120307}, DOI={10.7567/jjap.57.120307}, number={12}, journal={Japanese Journal of Applied Physics}, publisher={IOP Publishing}, author={Tsunegi, Sumito and Taniguchi, Tomohiro and Miwa, Shinji and Nakajima, Kohei and Yakushiji, Kay and Fukushima, Akio and Yuasa, Shinji and Kubota, Hitoshi}, year={2018}, month=oct, pages={120307} }

@article{Stieg_2011, title={Emergent Criticality in Complex Turing B‐Type Atomic Switch Networks}, volume={24}, ISSN={1521-4095}, url={http://dx.doi.org/10.1002/adma.201103053}, DOI={10.1002/adma.201103053}, number={2}, journal={Advanced Materials}, publisher={Wiley}, author={Stieg, Adam Z. and Avizienis, Audrius V. and Sillin, Henry O. and Martin‐Olmos, Cristina and Aono, Masakazu and Gimzewski, James K.}, year={2011}, month=oct, pages={286–293} }

@article{Nicola_2017, title={Supervised learning in spiking neural networks with FORCE training}, volume={8}, ISSN={2041-1723}, url={http://dx.doi.org/10.1038/s41467-017-01827-3}, DOI={10.1038/s41467-017-01827-3}, number={1}, journal={Nature Communications}, publisher={Springer Science and Business Media LLC}, author={Nicola, Wilten and Clopath, Claudia}, year={2017}, month=dec }

@article{Tanaka_2019, title={Recent advances in physical reservoir computing: A review}, volume={115}, ISSN={0893-6080}, url={http://dx.doi.org/10.1016/j.neunet.2019.03.005}, DOI={10.1016/j.neunet.2019.03.005}, journal={Neural Networks}, publisher={Elsevier BV}, author={Tanaka, Gouhei and Yamane, Toshiyuki and Héroux, Jean Benoit and Nakane, Ryosho and Kanazawa, Naoki and Takeda, Seiji and Numata, Hidetoshi and Nakano, Daiju and Hirose, Akira}, year={2019}, month=jul, pages={100–123} }

@article{Eliasmith_2012, title={A Large-Scale Model of the Functioning Brain}, volume={338}, ISSN={1095-9203}, url={http://dx.doi.org/10.1126/science.1225266}, DOI={10.1126/science.1225266}, number={6111}, journal={Science}, publisher={American Association for the Advancement of Science (AAAS)}, author={Eliasmith, Chris and Stewart, Terrence C. and Choo, Xuan and Bekolay, Trevor and DeWolf, Travis and Tang, Yichuan and Rasmussen, Daniel}, year={2012}, month=nov, pages={1202–1205} }

@article{Stewart_2012, title={Learning to Select Actions with Spiking Neurons in the Basal Ganglia}, volume={6}, ISSN={1662-4548}, url={http://dx.doi.org/10.3389/fnins.2012.00002}, DOI={10.3389/fnins.2012.00002}, journal={Frontiers in Neuroscience}, publisher={Frontiers Media SA}, author={Stewart, Terrence C. and Bekolay, Trevor and Eliasmith, Chris}, year={2012} }

@article{von_Neumann_1993, title={First draft of a report on the EDVAC}, volume={15}, ISSN={1058-6180}, url={http://dx.doi.org/10.1109/85.238389}, DOI={10.1109/85.238389}, number={4}, journal={IEEE Annals of the History of Computing}, publisher={Institute of Electrical and Electronics Engineers (IEEE)}, author={von Neumann, J.}, year={1993}, pages={27–75} }

@misc{motamedi_2023,
  doi = {10.48550/ARXIV.2304.03462},
  url = {https://arxiv.org/abs/2304.03462},
  author = {Motamedi, Arsalan and Zadeh-Haghighi, Hadi and Simon, Christoph},
  title = {Correlations Between Quantumness and Learning Performance in Reservoir Computing with a Single Oscillator},
  publisher = {arXiv},
  year = {2023},
  copyright = {arXiv.org perpetual, non-exclusive license}
}

@article{Govia_2021, title={Quantum reservoir computing with a single nonlinear oscillator}, volume={3}, ISSN={2643-1564}, url={http://dx.doi.org/10.1103/PhysRevResearch.3.013077}, DOI={10.1103/physrevresearch.3.013077}, number={1}, journal={Physical Review Research}, publisher={American Physical Society (APS)}, author={Govia, L. C. G. and Ribeill, G. J. and Rowlands, G. E. and Krovi, H. K. and Ohki, T. A.}, year={2021}, month=jan }

@article{Luchnikov_2019, title={Simulation Complexity of Open Quantum Dynamics: Connection with Tensor Networks}, volume={122}, ISSN={1079-7114}, url={http://dx.doi.org/10.1103/PhysRevLett.122.160401}, DOI={10.1103/physrevlett.122.160401}, number={16}, journal={Physical Review Letters}, publisher={American Physical Society (APS)}, author={Luchnikov, I. A. and Vintskevich, S. V. and Ouerdane, H. and Filippov, S. N.}, year={2019}, month=apr }

@article{Nokkala_2021, title={Gaussian states of continuous-variable quantum systems provide universal and versatile reservoir computing}, volume={4}, ISSN={2399-3650}, url={http://dx.doi.org/10.1038/s42005-021-00556-w}, DOI={10.1038/s42005-021-00556-w}, number={1}, journal={Communications Physics}, publisher={Springer Science and Business Media LLC}, author={Nokkala, Johannes and Martínez-Peña, Rodrigo and Giorgi, Gian Luca and Parigi, Valentina and Soriano, Miguel C. and Zambrini, Roberta}, year={2021}, month=mar }

@article{Garc_2023, title={Scalable Photonic Platform for Real-Time Quantum Reservoir Computing}, volume={20}, ISSN={2331-7019}, url={http://dx.doi.org/10.1103/PhysRevApplied.20.014051}, DOI={10.1103/physrevapplied.20.014051}, number={1}, journal={Physical Review Applied}, publisher={American Physical Society (APS)}, author={García-Beni, Jorge and Giorgi, Gian Luca and Soriano, Miguel C. and Zambrini, Roberta}, year={2023}, month=jul }

@article{Mujal_2023, title={Time-series quantum reservoir computing with weak and projective measurements}, volume={9}, ISSN={2056-6387}, url={http://dx.doi.org/10.1038/s41534-023-00682-z}, DOI={10.1038/s41534-023-00682-z}, number={1}, journal={npj Quantum Information}, publisher={Springer Science and Business Media LLC}, author={Mujal, Pere and Martínez-Peña, Rodrigo and Giorgi, Gian Luca and Soriano, Miguel C. and Zambrini, Roberta}, year={2023}, month=feb }

@article{Dudas_2023, title={Quantum reservoir computing implementation on coherently coupled quantum oscillators}, volume={9}, ISSN={2056-6387}, url={http://dx.doi.org/10.1038/s41534-023-00734-4}, DOI={10.1038/s41534-023-00734-4}, number={1}, journal={npj Quantum Information}, publisher={Springer Science and Business Media LLC}, author={Dudas, Julien and Carles, Baptiste and Plouet, Erwan and Mizrahi, Frank Alice and Grollier, Julie and Marković, Danijela}, year={2023}, month=jul }

@article{kobayashi_2024,
  author    = {Kaito Kobayashi and Keisuke Fujii and Naoki Yamamoto},
  title     = {Feedback-Driven Quantum Reservoir Computing for Time-Series Analysis},
  journal   = {PRX Quantum},
  volume    = {5},
  number    = {4},
  pages     = {040325},
  year      = {2024},
  month     = {November},
  doi       = {10.1103/PRXQuantum.5.040325},
  url       = {https://journals.aps.org/prxquantum/abstract/10.1103/PRXQuantum.5.040325}
}

@article{monomi_2025,
  author    = {Tomoya Monomi and Wataru Setoyama and Yoshihiko Hasegawa},
  title     = {Feedback-Enhanced Quantum Reservoir Computing with Weak Measurements},
  journal   = {arXiv preprint arXiv:2503.17939},
  year      = {2025},
  month     = {March},
  url       = {https://arxiv.org/abs/2503.17939}
}

@article{cindrak_2024,
  author    = {Saud {\v{C}}indrak and Brecht Donvil and Kathy L{\"u}dge and Lina Jaurigue},
  title     = {Enhancing the performance of quantum reservoir computing and solving the time-complexity problem by artificial memory restriction},
  journal   = {Physical Review Research},
  volume    = {6},
  number    = {1},
  pages     = {013051},
  year      = {2024},
  month     = {January},
  doi       = {10.1103/PhysRevResearch.6.013051},
  url       = {https://journals.aps.org/prresearch/abstract/10.1103/PhysRevResearch.6.013051}
}

@article{hamazaki_2022,
  author       = {Ryusuke Hamazaki and Masaya Nakagawa and Taiki Haga and Masahito Ueda},
  title        = {Lindbladian Many-Body Localization},
  journal      = {arXiv preprint arXiv:2206.02984},
  year         = {2022},
  month        = {June},
  eprint       = {2206.02984},
  archivePrefix= {arXiv},
  primaryClass = {cond-mat.dis-nn},
  url          = {https://arxiv.org/abs/2206.02984},
  doi          = {10.48550/arXiv.2206.02984}
}

@article{Nakajima2019,
  doi = {10.1103/physrevapplied.11.034021},
  url = {https://doi.org/10.1103/physrevapplied.11.034021},
  year = {2019},
  month = mar,
  publisher = {American Physical Society ({APS})},
  volume = {11},
  number = {3},
  pages = {034021-1},
  author = {Kohei Nakajima and Keisuke Fujii and Makoto Negoro and Kosuke Mitarai and Masahiro Kitagawa},
  title = {Boosting Computational Power through Spatial Multiplexing in Quantum Reservoir Computing},
  journal = {Physical Review Applied}
}

@article{fujii_2017,
  title = {Harnessing Disordered-Ensemble Quantum Dynamics for Machine Learning},
  author = {Fujii, Keisuke and Nakajima, Kohei},
  journal = {Phys. Rev. Appl.},
  volume = {8},
  issue = {2},
  pages = {024030},
  numpages = {20},
  year = {2017},
  month = {Aug},
  publisher = {American Physical Society},
  doi = {10.1103/PhysRevApplied.8.024030},
  url = {https://link.aps.org/doi/10.1103/PhysRevApplied.8.024030}
}

@article{martinez_2021,
  title = {Dynamical Phase Transitions in Quantum Reservoir Computing},
  author = {Mart\'{\i}nez-Pe\~na, Rodrigo and Giorgi, Gian Luca and Nokkala, Johannes and Soriano, Miguel C. and Zambrini, Roberta},
  journal = {Phys. Rev. Lett.},
  volume = {127},
  issue = {10},
  pages = {100502},
  numpages = {7},
  year = {2021},
  month = {Aug},
  publisher = {American Physical Society},
  doi = {10.1103/PhysRevLett.127.100502},
  url = {https://link.aps.org/doi/10.1103/PhysRevLett.127.100502}
}

@article{gotting_2023,
  title = {Exploring quantumness in quantum reservoir computing},
  author = {G\"otting, Niclas and Lohof, Frederik and Gies, Christopher},
  journal = {Phys. Rev. A},
  volume = {108},
  issue = {5},
  pages = {052427},
  numpages = {9},
  year = {2023},
  month = {Nov},
  publisher = {American Physical Society},
  doi = {10.1103/PhysRevA.108.052427},
  url = {https://link.aps.org/doi/10.1103/PhysRevA.108.052427}
}

@article{martinez_2020,
  title={Information Processing Capacity of Spin-Based Quantum Reservoir Computing Systems},
  author={Mart{\'i}nez-Pe{\~n}a, R. and Nokkala, J. and Giorgi, G. L. and Zambrini, R. and Soriano, M. C.},
  journal={Cognitive Computation},
  volume={15},
  number={},
  pages={1440--1451},
  year={2023},
  publisher={Springer},
  doi={10.1007/s12559-020-09772-y},
  url={https://link.springer.com/article/10.1007/s12559-020-09772-y}
}

@article{pfeuty_1971,
  title={The Ising model with a transverse field. II. Ground state properties},
  author={Pfeuty, P. and Elliott, R. J.},
  journal={Journal of Physics C: Solid State Physics},
  volume={4},
  pages={2370},
  doi={10.1088/0022-3719/4/15/024},
  year={1971}}

@article{stinchcombe_1973,
  title={Ising model in a transverse field. I. Basic theory},
  author={Stinchcombe, R. B.},
  journal={Journal of Physics C: Solid State Physics},
  volume={6},
  number={15},
  pages={2459},
  year={1973},
doi={10.1088/0022-3719/6/15/009},
  publisher={IOP Publishing Ltd}
}

@article{jaeger_2004,
  title={Harnessing Nonlinearity: Predicting Chaotic Systems and Saving Energy in Wireless Communication},
  author={Jaeger, Herbert and Haas, Harald},
  journal={Science},
  volume={304},
  pages={78--80},
  year={2004},
  publisher={American Association for the Advancement of Science},
  doi={10.1126/science.1091277},
  url={https://www.science.org/doi/10.1126/science.1091277}
}

@article{maass_2002,
  title={Real-Time Computing Without Stable States: A New Framework for Neural Computation Based on Perturbations},
  author={Maass, Wolfgang and Natschl{\"a}ger, Thomas and Markram, Henry},
  journal={Neural Computation},
  volume={14},
  number={11},
  pages={2531--2560},
  year={2002},
  publisher={MIT Press},
  doi={10.1162/089976602760407955},
  url={https://doi.org/10.1162/089976602760407955}}

@article{verstraeten_2007,
  title={An experimental unification of reservoir computing methods},
  author={Verstraeten, D. and Schrauwen, B. and D’Haene, M. and Stroobandt, D.},
  journal={Neural Networks},
  volume={20},
  issue={3},
  pages={391--403},
  year={2007},
  issn={0893-6080},
  doi={10.1016/j.neunet.2007.04.003},
  url={https://www.sciencedirect.com/science/article/pii/S089360800700038X}
}

@incollection{fujii_2021,
  title={Quantum Reservoir Computing: A Reservoir Approach Toward Quantum Machine Learning on Near-Term Quantum Devices},
  author={Fujii, K. and Nakajima, K.},
  booktitle={Reservoir Computing},
  editor={Nakajima, K. and Fischer, I.},
  series={Natural Computing Series},
  publisher={Springer},
  address={Singapore},
  year={2021},
  doi={10.1007/978-981-13-1687-6_18},
  url={https://doi.org/10.1007/978-981-13-1687-6_18}
}

@article{mujal_2021,
  title={Analytical evidence of nonlinearity in qubits and continuous-variable quantum reservoir computing},
  author={Mujal, Pere and Nokkala, Johannes and Mart{\'i}nez-Pe{\~n}a, Rodrigo and Giorgi, Gian Luca and Soriano, Miguel C and Zambrini, Roberta},
  journal={J. of Phys. Complex.},
  volume={2},
  number={4},
  year={2021},
  publisher={IOP Publishing Ltd},
  doi={10.1088/2632-072X/ac340e},
  pages={045008}
}

@article{vidal_2002,
  title = {Computable measure of entanglement},
  author = {Vidal, G. and Werner, R. F.},
  journal = {Phys. Rev. A},
  volume = {65},
  issue = {3},
  pages = {032314},
  numpages = {11},
  year = {2002},
  month = {Feb},
  publisher = {American Physical Society},
  doi = {10.1103/PhysRevA.65.032314},
  url = {https://link.aps.org/doi/10.1103/PhysRevA.65.032314}
}

@article{plenio_2005,
  title = {Logarithmic Negativity: A Full Entanglement Monotone That is not Convex},
  author = {Plenio, M. B.},
  journal = {Phys. Rev. Lett.},
  volume = {95},
  issue = {9},
  pages = {090503},
  numpages = {4},
  year = {2005},
  month = {Aug},
  publisher = {American Physical Society},
  doi = {10.1103/PhysRevLett.95.090503},
  url = {https://link.aps.org/doi/10.1103/PhysRevLett.95.090503}
}

@article{araiza_2022,
  title = {Quantum Reservoir Computing Using Arrays of Rydberg Atoms},
  author = {Bravo, Rodrigo Araiza and Najafi, Khadijeh and Gao, Xun and Yelin, Susanne F.},
  journal = {PRX Quantum},
  volume = {3},
  issue = {3},
  pages = {030325},
  numpages = {19},
  year = {2022},
  month = {Aug},
  publisher = {American Physical Society},
  doi = {10.1103/PRXQuantum.3.030325},
  url = {https://link.aps.org/doi/10.1103/PRXQuantum.3.030325}
}

@article{sannia_2024,
  doi       = {10.22331/q-2024-03-20-1291},
  url       = {https://doi.org/10.22331/q-2024-03-20-1291},
  title     = {Dissipation as a resource for {Q}uantum {R}eservoir {C}omputing},
  author    = {Sannia, Antonio and Mart{\'{\i}}nez-Pe{\~{n}}a, Rodrigo and Soriano, Miguel C. and Giorgi, Gian Luca and Zambrini, Roberta},
  journal   = {Quantum},
  issn      = {2521-327X},
  publisher = {{Verein zur F{\"{o}}rderung des Open Access Publizierens in den Quantenwissenschaften}},
  volume    = {8},
  pages     = {1291},
  month     = mar,
  year      = {2024}
}

@article{franceschetto_2024,
  author       = {Giacomo Franceschetto and Marcin Płodzień and Maciej Lewenstein and Antonio Acín and Pere Mujal},
  title        = {Harnessing quantum back-action for time-series processing},
  journal      = {arXiv preprint arXiv:2411.03979},
  year         = {2024},
  month        = {November},
  eprint       = {2411.03979},
  archivePrefix= {arXiv},
  primaryClass = {quant-ph},
  doi          = {10.48550/arXiv.2411.03979},
  url          = {https://arxiv.org/abs/2411.03979}
}

@article{gu_2025,
  title = {Magic-Induced Computational Separation in Entanglement Theory},
  author = {Gu, Andi and Oliviero, Salvatore F.E. and Leone, Lorenzo},
  journal = {PRX Quantum},
  volume = {6},
  issue = {2},
  pages = {020324},
  numpages = {47},
  year = {2025},
  month = {May},
  publisher = {American Physical Society},
  doi = {10.1103/PRXQuantum.6.020324},
  url = {https://link.aps.org/doi/10.1103/PRXQuantum.6.020324}
}

@article{leone_2022,
  title = {Stabilizer R\'enyi Entropy},
  author = {Leone, Lorenzo and Oliviero, Salvatore F. E. and Hamma, Alioscia},
  journal = {Phys. Rev. Lett.},
  volume = {128},
  issue = {5},
  pages = {050402},
  numpages = {5},
  year = {2022},
  month = {Feb},
  publisher = {American Physical Society},
  doi = {10.1103/PhysRevLett.128.050402},
  url = {https://link.aps.org/doi/10.1103/PhysRevLett.128.050402}
}

@misc{maronese_2025,
      title={High-expressibility Quantum Neural Networks using only classical resources}, 
      author={Marco Maronese and Francesco Ferrari and Matteo Vandelli and Daniele Dragoni},
      year={2025},
      eprint={2506.13605},
      archivePrefix={arXiv},
      primaryClass={quant-ph},
      url={https://arxiv.org/abs/2506.13605}, 
}

@article{garcia_2024,
author = {Jorge Garc\'{i}a-Beni and Gian Luca Giorgi and Miguel C. Soriano and Roberta Zambrini},
journal = {Opt. Express},
number = {4},
pages = {6733--6747},
publisher = {Optica Publishing Group},
title = {Squeezing as a resource for time series processing in quantum reservoir computing},
volume = {32},
month = {Feb},
year = {2024},
url = {https://opg.optica.org/oe/abstract.cfm?URI=oe-32-4-6733},
doi = {10.1364/OE.507684},
}

@article{kora_2024,
  title = {Frequency- and dissipation-dependent entanglement advantage in spin-network quantum reservoir computing},
  author = {Kora, Youssef and Zadeh-Haghighi, Hadi and Stewart, Terrence C. and Heshami, Khabat and Simon, Christoph},
  journal = {Phys. Rev. A},
  volume = {110},
  issue = {4},
  pages = {042416},
  numpages = {11},
  year = {2024},
  month = {Oct},
  publisher = {American Physical Society},
  doi = {10.1103/PhysRevA.110.042416},
  url = {https://link.aps.org/doi/10.1103/PhysRevA.110.042416}
}

@article{fangjun_2023,
  title = {Tackling Sampling Noise in Physical Systems for Machine Learning Applications: Fundamental Limits and Eigentasks},
  author = {Hu, Fangjun and Angelatos, Gerasimos and Khan, Saeed A. and Vives, Marti and T\"ureci, Esin and Bello, Leon and Rowlands, Graham E. and Ribeill, Guilhem J. and T\"ureci, Hakan E.},
  journal = {Phys. Rev. X},
  volume = {13},
  issue = {4},
  pages = {041020},
  numpages = {34},
  year = {2023},
  month = {Oct},
  publisher = {American Physical Society},
}

@article{palacios_2024,
  author    = {Ana Palacios and Rodrigo Mart{\'i}nez-Pe{\~n}a and Miguel C. Soriano and Gian Luca Giorgi and Roberta Zambrini},
  title     = {Role of coherence in many-body Quantum Reservoir Computing},
  journal   = {Communications Physics},
  volume    = {7},
  number    = {369},
  year      = {2024},
  month     = {November},
  doi       = {10.1038/s42005-024-01859-4},
  url       = {https://www.nature.com/articles/s42005-024-01859-4}
}

@article{streltsov_2017,
  author    = {Alexander Streltsov and Gerardo Adesso and Martin B. Plenio},
  title     = {Colloquium: Quantum coherence as a resource},
  journal   = {Reviews of Modern Physics},
  volume    = {89},
  number    = {4},
  pages     = {041003},
  year      = {2017},
  month     = {October},
  doi       = {10.1103/RevModPhys.89.041003},
  url       = {https://journals.aps.org/rmp/abstract/10.1103/RevModPhys.89.041003}
}

@article{ehlers_2025,
  author    = {Peter J. Ehlers and Hendra I. Nurdin and Daniel Soh},
  title     = {Stochastic reservoir computers},
  journal   = {Nature Communications},
  volume    = {16},
  article   = {3070},
  year      = {2025},
  month     = {March},
  doi       = {10.1038/s41467-025-58349-6},
  url       = {https://www.nature.com/articles/s41467-025-58349-6}
}

@article{baumgratz_2014,
  author    = {T. Baumgratz and M. Cramer and M. B. Plenio},
  title     = {Quantifying Coherence},
  journal   = {Physical Review Letters},
  volume    = {113},
  number    = {14},
  pages     = {140401},
  year      = {2014},
  month     = {October},
  doi       = {10.1103/PhysRevLett.113.140401},
  url       = {https://journals.aps.org/prl/abstract/10.1103/PhysRevLett.113.140401}
}

@article{suzuki_2022,
  title={Natural quantum reservoir computing for temporal information processing},
  author={Suzuki, Y. and Gao, Q. and Pradel, K.C. and others},
  journal={Scientific Reports},
  volume={12},
  pages={1353},
  year={2022},
  publisher={Nature Publishing Group},
  doi={10.1038/s41598-022-05061-w},
  url={https://doi.org/10.1038/s41598-022-05061-w}
}

@misc{gilboa_2023,
      title={Exponential Quantum Communication Advantage in Distributed Learning}, 
      author={Dar Gilboa and Jarrod R. McClean},
      year={2023},
      eprint={2310.07136},
      archivePrefix={arXiv},
      primaryClass={quant-ph}
}

@article{dambre_2012,
  title={Information Processing Capacity of Dynamical Systems},
  author={Dambre, Joni and Verstraeten, David and Schrauwen, Benjamin and Massar, Serge},
  journal={Scientific Reports},
  volume={2},
  pages={514},
  year={2012},
  publisher={Nature Publishing Group},
  doi={10.1038/srep00514},
  url={https://doi.org/10.1038/srep00514}
}

@book{breuer_2002,
  title={The Theory of Open Quantum Systems},
  author={Breuer, H. and Petruccione, F.},
  year={2002},
  publisher={Oxford University Press},
  address={Oxford}
}

@article{carroll_2022,
  title={Optimizing memory in reservoir computers},
  author={Carroll, T. L.},
  journal={Chaos},
  volume={32},
  pages={023123},
  year={2022},
  doi={https://doi.org/10.1063/5.0078151},
}


\end{document}